\documentclass[prl,twocolumn,showpacs,superscriptaddress]{revtex4}

\usepackage{graphicx}

\def\nbOne{{\mathchoice {\rm 1\mskip-4mu l} {\rm 1\mskip-4mu l}
{\rm 1\mskip-4.5mu l} {\rm 1\mskip-5mu l}}}
\begin{document}

\title{Entanglement entropy in the Lipkin-Meshkov-Glick model}

\author{Jos\'e I. Latorre}
\author{Rom\'an Or\'us}
\author{Enrique Rico}
\affiliation{Dept. d'Estructura i Constituents de la Mat\`eria,
   Univ. Barcelona, 08028, Barcelona, Spain}

\author{Julien Vidal}
\affiliation{Laboratoire de Physique Th\'eorique de la Mati\`ere Condens\'ee, CNRS UMR 7600,
Universit\'e Pierre et Marie Curie, 4 Place Jussieu, 75252 Paris Cedex 05, France}

\begin{abstract}

We analyze the entanglement entropy in the Lipkin-Meshkov-Glick model, which describes mutually interacting spins half embedded in a magnetic field.  This entropy displays a singularity at the critical point that we study as a function of the interaction anisotropy, the magnetic field,  and the system size. Results emerging from our analysis are surprisingly similar to those found for the one-dimensional $XY$
chain.

\end{abstract}


\maketitle

%
%

Within the last few years, entanglement properties of spin systems
have attracted much attention. As initially shown in one-dimensional $(1D)$
spin chains \cite{Osterloh02,Osborne02,Latorre03,Latorre04_1},
observables measuring this genuine quantum mechanical feature are 
strongly affected by the existence of a quantum phase
transition. For instance, the so-called concurrence \cite{Wootters98}
that quantifies the two-spin entanglement displays some nontrivial
universal scaling properties.  Similarly, the von Neumann entropy
which rather characterizes the bipartite entanglement between any two
subsystems scales logarithmically with the typical size $L$ of these
subsystems at the quantum critical point, with a prefactor given by
the central charge of the corresponding theory
\cite{Latorre03,Latorre04_1,Jin04,Refael04}. Note that the role played
by the boundaries in these conformal invariant systems has been
only recently elucidated \cite{Calabrese04}.  Apart from $1D$
systems, very few models have been studied so far
\cite{Syljuasen03_2,Hamma05,Latorre04_2,Orus04,Plenio04}, either due to the
absence of exact solution or to a difficult numerical treatment.
In this context, the Lipkin-Meshkov-Glick (LMG) model \cite{Lipkin65,Meshkov65,Glick65} 
discussed here has drawn much attention since it allows for very efficient numerical treatment as
well as analytical calculations.  Introduced by Lipkin, Meshkov
and Glick in Nuclear Physics, this model has been the
subject of intensive studies during the last two decades because of
its relevance for quantum tunneling of bosons between two levels.  It
is thus of prime interest to describe in particular the Josephson
effect in two-mode Bose-Einstein condensates \cite{Milburn97,Cirac98}.
The entanglement properties of this model have been already discussed through the concurrence, which exhibits a cusp-like behavior at the critical point \cite{Vidal04_1,Vidal04_2,Dusuel04_3,Dusuel04_4} as
well as interesting dynamical properties \cite{Vidal04_3}. Note that simi\-lar results have also been
obtained in the Dicke model \cite{Dicke54,Lambert04} which can be mapped onto the LMG model in some cases \cite{Reslen04}, or in the reduced BCS model \cite{Dusuel05}.

In this letter, we analyze the von Neumann entropy computed from the ground state of the LMG
model.  We show that, at the critical point, it behaves logarithmically with the size of the blocks $L$ used
in the bipartite decomposition of the density matrix with a prefactor that depends on the anisotropy
parameter tuning the universality class. We also discuss the dependence of the entropy with the
magnetic field and stress the analogy with $1D$ systems.
%
%

The LMG model is defined by the Hamiltonian
%
%
\begin{equation}
H=-\frac{\lambda}{N} \sum_{i<j} \left( \sigma_x^i \sigma_x^j + \gamma
\sigma_y^i \sigma_y^j
\right) - h \sum_i \sigma_z^i,
\label{hamil0}
\end{equation}
%
%
where $\sigma_{\alpha}^k$ is the Pauli matrix at position
$k$ in the direction $\alpha$, and $N$ the total number of spins. This
Hamiltonian describes a set of spins half located at the vertices of a
$N$-dimensional simplex (complete graph) interacting via a
ferromagnetic coupling ($\lambda>0$) in the $xy$-spin plane, $\gamma$
being an anisotropy parameter and $h$ an external magnetic field
applied along the $z$ direction.  The Hamiltonian (\ref{hamil0}) can
be written in terms of the total spin operators $S_{\alpha}= \sum_i
\sigma_{\alpha}^i / 2$ as
%
%
\begin{eqnarray}
H&=& - {\lambda \over N} (1+\gamma) \left({\bf S}^2-S_z^2 -N/2 \right)
-2h S_z  \nonumber \\
&&- {\lambda \over 2 N} (1-\gamma)\left(S_+^2+S_-^2\right) \ . \label{hamil}
\end{eqnarray}
%
%
In the following, for simplicity we set  $\lambda=1$, and since the spectrum of $H$ is even
under the transformation $h \leftrightarrow -h$ \cite{Vidal04_3}, we restrict our
analysis to the region $h\geq 0$. Furthermore, we only consider the maximum spin sector $S=N/2$, to
which the ground state is known to belong.
A convenient basis of this subspace is spanned by the so-called Dicke states 
$|N/2,M \rangle$, which are fully symmetric under the permutation group and are eigenstates of
${\bf S}^2$ and $S_z$, with eigenvalues $N(N+2)/4$ and $M$, respectively.

In order to study the entanglement of the ground
state, we need to define an appropriate figure of
merit. Following the ideas presented in
\cite{Latorre03,Latorre04_1,Latorre04_2,Orus04}, we consider the von Neumann entropy
associated to the ground-state reduced density matrix $\rho_{L,N}$ of
a block of size $L$ out of the total $N$ spins, $S_{L,N}= -{\rm
Tr}\: (\rho_{L,N} \log_2 \rho_{L,N})$ and analyze its behavior as $L$ is
changed, both keeping $N$ finite or sending it to infinity.  Note that since the ground
state reduced density matrix is spanned by the set of $(L+1)$
Dicke states, the entropy of entanglement obeys the constraint $S_{L,N} \le
{\rm log}_2 (L+1)$ for all $L$ and $N$, where the upper bound corresponds to
the entropy of the maximally mixed state $\rho_{L,N} = \nbOne/(L+1)$ in
the Dicke basis. This
argument implies that entanglement, as measured by the Von Neumann
entropy, cannot grow faster than the typical logarithmic scaling
observed in $1D$ quantum spin chains
\cite{Latorre03,Latorre04_1}. 

%
%

\begin{figure}
\centering
\includegraphics[angle=-90, width=0.55\textwidth]{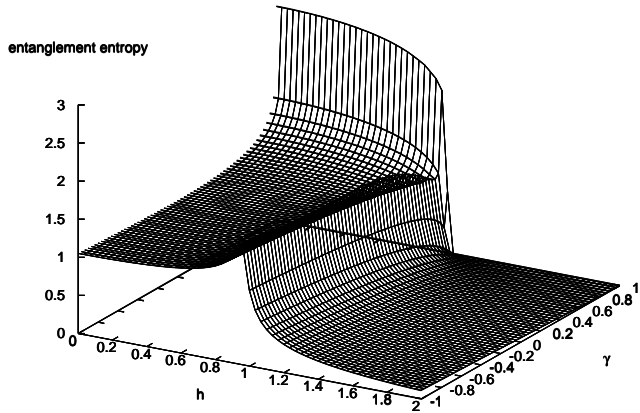}
\caption{Entanglement entropy for $N=500$ and $L=125$ as a function of $h$ and $\gamma$.} 
\label{fig:Shg500}
\end{figure}

In order to study the different entanglement regimes, we compute the entropy in the
plane spanned by $\gamma$ and $h$.  This numerical computation can be laid out
analytically taking advantage of the Hamiltonian symmetries to reduce
the complexity of the numerical task to a polynomial growth in $N$. Results are displayed in
Fig. \ref{fig:Shg500} for $N=500$ and $L=125$. For $\gamma \ne 1$, one clearly
observes an anomaly at the critical point $h=1$, whereas the entropy goes to
zero at large $h$ since the ground state is then a fully polarized state in
the field direction. In the zero field
limit, the entropy saturates when the
size of the system increases and goes to $S_{L,N}=1$ for $\gamma = 0$, where the ground
state approaches a GHZ-like (cat-like) state as in the
Ising spin chain \cite{Latorre03,Latorre04_1}. By contrast, for $\gamma = 1$,
the entropy increases with the size of the system in the region $0 \le h < 1$
and jumps to zero at $h=1$ as we shall now discuss.

%
%

In the isotropic case ($\gamma=1$), it is possible to
analytically compute the entropy of entanglement since, at this point, $H$ is
diagonal in the Dicke basis.  The
ground-state energy is given by $E_0 (h,\gamma=1) = 
-{N \over 2} + \frac{2}{N} M^2  -2 h M $, with 
%
%
\begin{equation}
M=\Bigg\{ 
\begin{array}{ccc}
I(h N /2), & {\rm for}&  0\leq h < 1 \\ 
N/2, & {\rm for}&  h \geq 1 
\end{array}
\ ,
\label{MGS}
\end{equation}
%
%
and the corresponding eigenvector is simply $|N/2,M\rangle$. Here, $I(x)$ denotes the round value of $x$. 

To calculate the entropy, it is convenient to introduce the number $n$ of
spins ``up" so that $M=n-N/2$, and to write this state in a bipartite
form. Indeed, since Dicke states are completely symmetric under any
permutation of sites, it is straightforward to see that the ground
state can be written as a sum of byproducts of Dicke states:
%
%
\begin{eqnarray}
\label{dicke}
|N/2,n-N/2\rangle &=&\sum_{l=0}^L p_l^{1/2} |L/2,l-L/2\rangle \otimes  \\
&&  |(N-L)/2 ,n-l-(N-L)/2\rangle \nonumber,
\end{eqnarray}
%
%
where the partition is made between two blocks of
size $L$ and $(N-L)$ and 
%
%
\begin{equation}
p_l = 
{\left(
\begin{array}{c}
L
\\
l
\end{array}
\right)
\left(
\begin{array}{c}
N-L
\\
n-l
\end{array}
\right)
\over
\left(
\begin{array}{c}
N
\\
n
\end{array}
\right)
}.
\end{equation}
%
%
The ground-state entropy is then simply given by $S_{L,N} (h,\gamma)=
-\sum_{l=0}^L p_l \log_2{p_l}$. 
In the limit  $N,  L \gg 1$, the hypergeometric distribution of the $p_l$ can be
recast into a Gaussian distribution $p_l \simeq p_l^g=\frac{1}{\sqrt{2 \pi} \sigma} \exp\left[-\frac{(l-\bar
l)^2}{2\sigma^2}\right]$, of mean value $\bar{l}=n {L \over N}$ and variance 
%
%
\begin{eqnarray}
\label{variance}
\sigma^2 = n(N-n)\frac{(N-L)L}{N^3},
\end{eqnarray}
%
%
where we have retained the subleading term in $(N-L)$ to explicitly
preserve the symmetry $S_{L,N}=S_{N-L,N}$. The entropy then reads
%
%
\begin{equation}
\label{gaussian}
- \int_{-\infty}^\infty\ {\rm d}l\  p_l^g  \log_2{p_l^g}
= \frac{1}{2} \left(\log_2{e}+ \log_2{2\pi} + \log_2{\sigma^2} \right) \ , 
\end{equation}
%
%
and only depends on its variance as expected for a Gaussian distribution. Let us mention that this result has been recently obtained in the context of the ferromagnetic Heisenberg chain \cite{Popkov05}.  Of course, for $h \ge 1$, the
entanglement entropy is exactly zero since the ground state is, in this case,
fully polarized in the magnetic field direction ($n=N$).  For $h \in [0,1[$ and in 
the limit $N, L \gg 1$, Eqs. (\ref{MGS}), (\ref{variance}) and (\ref{gaussian}) lead to
%
%
\begin{equation}
S_{L,N} (h,\gamma=1) \sim \frac{1}{2} \log_2{[L(N-L)/N ]}.
\label{eq:entropy_iso}
\end{equation}
%
%

Moreover, the  dependence of the entropy with the magnetic field is given by
%
%
\begin{equation}
S_{L,N} (h,\gamma=1) -S_{L,N} (h=0,\gamma=1) \sim \frac{1}{2} \log_2 {\left(1-h^2 \right)},
\label{eq:Shiso}
\end{equation}
%
and thus diverges, at fixed $L$ and $N$, in the limit $h\rightarrow 1^-$.

%
 %
%

Let us now discuss the more general situation $\gamma \ne 1$, for which no
simple analytical solution exists. In this case, the ground state is a
superposition of Dicke states with coefficients that can be easily
determined by numerical diagonalizations. Upon tracing out $(N-L)$
spins, each Dicke state decomposes as in Eq. (\ref{dicke}). 
It is then easy to build the $(L+1)\times(L+1)$ ground-state reduced density
matrix and to compute its associated entropy.  

%
%
\begin{figure}[h]
\centering
\includegraphics[angle=-90, width=0.5\textwidth]{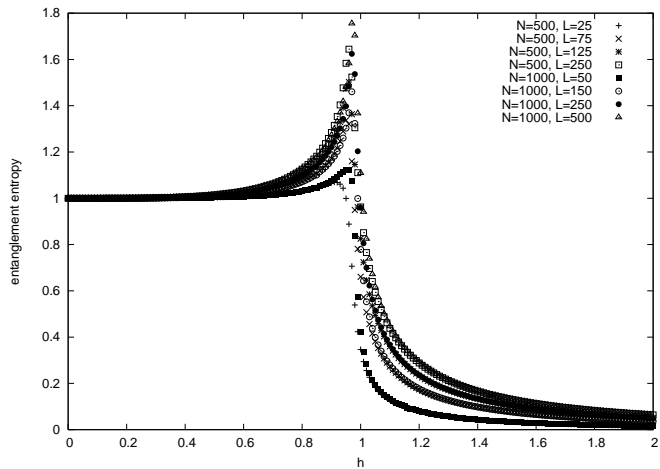}
\caption{Entanglement entropy at $\gamma = 0$ as a function of $h$ for
  different values of $N$ and $L$. Outside of the critical region, the entropy only depends on the 
 ratio $L/N$.}  
\label{fig:Shg0N}
\end{figure}
%
%

\begin{figure}[h]
\centering
\includegraphics[angle=-90, width=0.5\textwidth]{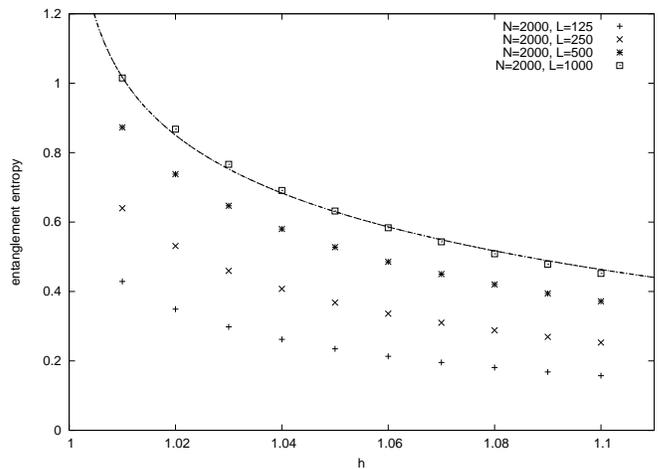}
\caption{Entanglement entropy as a function of $h$ near the critical point for $\gamma=0$. The solid line corresponds to the fitting law 
(\ref{eq:Shani}) with $a=1/6$.}
\label{fig:Shg02000}
\end{figure}

\begin{figure}[h]
\centering
\includegraphics[angle=-90, width=0.5\textwidth]{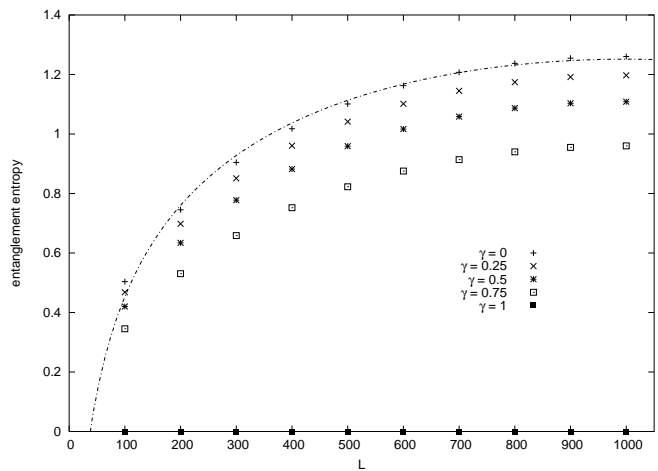}
\caption{Entanglement entropy as a function of $L$ at the critical point for 
different  $\gamma$ and $N=2000$. The solid line corresponds to the fitting law 
(\ref{eq:entropy_ani}) with $b=1/3$.}
\label{fig:Sh1L2000}
\end{figure}

\begin{figure}[h]
\centering
\includegraphics[angle=-90, width=0.5\textwidth]{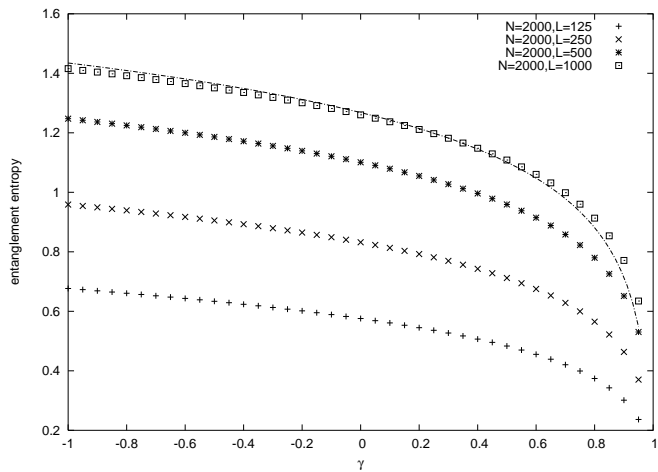}
\caption{Entanglement entropy at the critical point $h=1$ as a 
function of $\gamma$. The solid line corresponds to the fitting law 
(\ref{eq:Sgamma}) with $d=1/6$.}
\label{fig:Sh1g2000}
\end{figure}

We have displayed in Fig. \ref{fig:Shg0N}, the behavior of the entropy as a
function of $h$, for different values of the ratio $L/N$ and for $\gamma=0$. 
For $h\neq 1$, the entropy only depends on the ratio $L/N$. For any $\gamma$, 
at fixed $L/N$ and in the limit $h\rightarrow \infty$, the entropy goes to
zero since the ground state becomes then fully polarized in the field
direction. Note that the entropy also vanishes, at $h>1$, in the limit 
$L/N \rightarrow 0$ where the entanglement properties are trivial. 
In the zero field limit, the entropy goes to a constant which depends on
$\gamma$ and equals 1 at $\gamma=0$ since the ground state is then a GHZ-like 
state made up of spins pointing in $\pm x$ directions. 
Close to criticality, the entropy displays a logarithmic divergence,  which we numerically found  to obey the law 
\begin{equation}
S_{L,N} (h,\gamma) \sim -a \log_2{|1-h|},
\label{eq:Shani}
\end{equation}
where $a$ is very close to 1/6 for $N,L \gg 1$ as can be seen in Fig. \ref{fig:Shg02000}.
%
%
%
%
At the critical point, the entropy has a 
nontrivial behavior that we have studied focusing on 
the point $\gamma=0$ which is representative of the class $\gamma \ne 1$. There, the entropy also scales logarithmically  with $L$ as in the isotropic case, but with a different prefactor. 
More precisely, we find
%
%
\begin{equation}
S_{L,N} (h=1,\gamma \neq 1) \sim b \log_2{[L(N-L)/N ]}.
\label{eq:entropy_ani}
\end{equation}
%
%
For the finite-size systems investigated here, the prefactor varies when 
either the ratio $L/N$ or $\gamma$ is changed, as  can be seen in  
Fig. \ref{fig:Sh1L2000}. However,  in the thermodynamical limit  $N, L \gg 1$ 
(and finite $L/N$), it is likely that $b=1/3$ and  does not depend on
$\gamma$.
%
%
%
%
In addition, at fixed $L$ and $N$, the entropy also depends on the anisotropy 
parameter logarithmically as
%
%
\begin{equation}
S_{L,N}(h=1,\gamma) - S_{L,N}(h=1,\gamma=0)\sim d  \log_2(1-\gamma),
\label{eq:Sgamma}
\end{equation}
%
%
for all $-1 \le \gamma < 1$, as can be seen in Fig. \ref{fig:Sh1g2000}.
Here again, it is likely that, in the thermodynamical limit, $d$  
has a simple (rational) value which, from our data, seems to be $1/6$.  
%
%
%
%
It is important to keep in mind that the limit $\gamma \rightarrow 1$ and the thermodynamical limit do not commute, so that Eq. (\ref{eq:Sgamma}) is only valid for $\gamma \neq 1$.
%
%
%

Let us now compare these results with those found in the $1D$ $XY$ model. 
As for the LMG model, the $XY$ chain  has two different universality classes 
depending on the anisotropy parameter. At the critical point, the entropy 
has been found to behave as \cite{Latorre03,Latorre04_1,Jin03}
%
%
\begin{equation}
S_{L,N} \sim {c \over 3} \log_2
[L(N-L)/N],
\label{scaling}
\end{equation}
%
%
where $c$ is the central charge of the corresponding conformal theory
\cite{Holzhey94}. For the isotropic case, the critical model is indeed
described by a free boson theory  with $c_{XX}=1$ whereas the anisotropic 
case corresponds to a free fermion theory with with $c_{XY}=1/2$.
It is striking to see that the entropy in the LMG model has the same 
logarithmic dependence with some prefactor which, as in the $1D$ case, depends only 
on the universality classes [see Eqs. (\ref{eq:entropy_iso}) and 
(\ref{eq:entropy_ani})].
Concerning the dependence with the magnetic field and with the anisotropy parameter,  it is also worth noting that logarithmic behaviors  (\ref{eq:Shiso}), (\ref{eq:Shani}), and  (\ref{eq:Sgamma}) are similar to those found in the $XX$ and $XY$ chain \cite{Its04} except that the prefactor in the LMG model are different.
On top,  the  behavior of this model with respect to
majorization for $\gamma \ne 1$ and as $h$ departs from its critical value is completely analogous 
to the $1D$ $XY$ model \cite{Latorre05_1}, namely, 
all the eigenvalues of the reduced density matrix obey 
a strict majorization relation as $h$ grows, while for decreasing $h$ one of these 
eigenvalues drives the system towards a GHZ-like state in such a way that
majorization is strictly obeyed in the thermodynamical limit. This behavior implies a 
very strong ``sense of order" in the ground state, in complete analogy to the
$XY$ chain.  

The logarithmic scaling with $L$ of the ground-state entanglement entropy we
have found for the LMG model, with well-defined values of the scaling
coefficients in the whole parameter space, entices the search for a precise
construction of an underlying one-dimensional local field theory. 
In order to further clarify this statement, we have explicitly analyzed the
distribution and degeneracy of eigenvalues of 
the reduced density matrix of the system, that is, the structure of the
partition function of the model. We have observed that, at
$\gamma = 1$, the spectrum is doubly-degenerate with the exception of the first eigenvalue
and follows a Gaussian behavior (as expected from the previous analytical
calculations), while at $\gamma = 0$ the spectrum is doubly degenerate and equally spaced in
logarithmic scale. The behavior of the spectrum for $\gamma = 0$ accommodates
to the typical structure imposed by the Virasoro algebra over the
highest-weight operators in conformal field theories \cite{Ginsbarg88}. 
This observation might be a hint of a possible conformal 
invariance underlying the LMG model in some regions of its parameter space.

%
%

Finally, let us mention that during the completion of this work, the
entanglement entropy has also been calculated for the antiferromagnetic 
LMG model \cite{Unanyan04} for which the ground state is known 
exactly \cite{Unanyan03,Vidal04_2}.\\
 
 \acknowledgments

We wish to thank C. A. L\"{u}tken and R. Mosseri for fruitful and valuable
discussions. We also acknowledge financial support from projects MECD AP2001-1676, MCYT
FPA2001-3598, GC2001SGR-00065 and IST-1999-11053.  


\end{document}